\documentclass[twocolumn,showpacs,preprintnumbers,superscriptaddress,amsmath,amssymb,aps,pre]{revtex4-1}
\usepackage{epsfig}
\usepackage{graphicx}
\usepackage{dcolumn}
\usepackage{bm}
\usepackage{times}
\usepackage{xcolor}
\usepackage{booktabs}

\begin{document}

\author{Liming Pan}
\affiliation{Web Sciences Center, School of Computer Science and Engineering, University of Electronic Science and Technology of China, Chengdu, 611731, China}

\author{Wei Wang}\email{wwzqbx@hotmail.com}

\affiliation{Cybersecurity Research Institute, Sichuan University, Chengdu 610065, China}
\affiliation{Web Sciences Center, School of Computer Science and Engineering, University of Electronic Science and Technology of China, Chengdu, 611731, China}

\author{Shimin Cai}
\affiliation{Web Sciences Center, School of Computer Science and Engineering, University of Electronic Science and Technology of China, Chengdu, 611731, China}

\author{Tao Zhou}
\affiliation{Web Sciences Center, School of Computer Science and Engineering, University of Electronic Science and Technology of China, Chengdu, 611731, China}

\date{\today}

\title{Optimal interlayer structure for promoting spreading of SIS model in two-layer networks}

\begin{abstract}
Real-world systems, ranging from social and biological to infrastructural, can be modeled by multilayer networks. Promoting spreading dynamics in multilayer networks may significantly
facilitate electronic advertising and predicting popular scientific publications. In this study, we propose a strategy for promoting the spreading dynamics of the susceptible-infected-susceptible model by adding one interconnecting edge between two isolated networks. By applying a perturbation method to the discrete Markovian chain approach, we derive an index that estimates the spreading prevalence in the interconnected network. The index can be interpreted as a variant of Katz centrality, where the adjacency matrix is replaced by a weighted matrix with weights depending on the dynamical information of the spreading process. Edges that are less infected at one end and its neighborhood but highly infected at the other will have larger weights. We verify the effectiveness of the proposed strategy on small networks by exhaustively examining all latent edges and demonstrate that performance is optimal or near-optimal. For large synthetic and real-world networks, the proposed method always outperforms other static strategies such as connecting nodes with the highest degree or eigenvector centrality.

\end{abstract}

\maketitle

\section{Introduction} \label{sec:intro}
Promoting spreading dynamics in networked systems is attracting considerable attention in network science, statistical physics, and computer science~\cite{lu2016vital}. Maximizing spreading prevalence is of both theoretical and practical importance for achieving better information spreading and providing vaccination guidance. Strategies for maximizing spreading prevalence can be roughly divided into three categories: identifying vital nodes~\cite{kitsak2010identification,
morone2015influence,lu2016h,chen2009efficient,chen2010scalable,
morone2016collective,hu2018local,ji2017effective,ren2014iterative,
liao2017ranking,pei2013spreading,chen2012identifying,liu2017accurate}, designing effective transmission strategies~\cite{yang2008optimal,gao2016effective,
yang2008selectivity,roshani2012effects,gao2017promoting,cui2018close}, and performing network structural perturbations~\cite{aguirre2013successful,del2018finding,
milanese2010approximating,van2010influence}. For vital node identification, centrality measures, such as K-core, H-index, betweenness, and degree centrality, are assigned to network nodes. Nodes with high centrality are then chosen to be initial seeds for spreading. For effective transmission, spreading protocols have been designed to avoid invalid contacts (i.e., contacts among infected nodes). For performing structural perturbations, networks are modified slightly to promote spreading~\cite{aguirre2013successful}. Structural perturbations are also widely applied to enhance network synchronizability ~\cite{aguirre2014synchronization,li2016synchronizability,
wei2018synchronizability,wei2018maximizing,dai2019interconnecting}.

The effectiveness of strategies for promoting spreading relies on the underlying spreading models. Various models, such as the susceptible-infected-susceptible (SIS), susceptible-infected-recovered (SIR), and threshold models, have been employed to test the effectiveness of such strategies~\cite{lu2016vital}. These spreading models can be divided into two classes, namely, simple and complex contagions~\cite{centola2018behavior,guilbeault2018complex}. In simple contagions, a susceptible individual could be infected by a single contact with an infected individual. Simple contagions are usually applied to model disease spreading~\cite{pastor2001epidemic} and simple information spreading (e.g., hashtags~\cite{romero2011differences}). In complex contagions, individuals evaluate the legitimacy of the information and make a risk assessment; the probability of infection increases with the cumulated number of contacts with other infected social peers. This mechanism is called social reinforcement ~\cite{watts2002simple,centola2007complex,aral2017exercise,lu2011small}. Complex contagions are usually applied to model complex information spreading (e.g., political information~\cite{romero2011differences}) and behavior adoption~\cite{centola2007complex,aral2017exercise,unicomb2019reentrant,karsai2014complex}.
More complex spreading mechanisms, such as the coevolution of multiple diseases and/or information, are discussed
in the recent review~\cite{wang2019coevolution}.

The spreading dynamics in multilayer networks can be fundamentally different from that in single-layer networks~\cite{da2018epidemic,de2016physics,granell2013dynamical,
granell2014competing,wang2014asymmetrically,chen2018optimal,
gao2012networks,tejedor2018diffusion,wang2018social}.
For instance, Granell \emph{et al.}~\cite{granell2013dynamical} demonstrated that epidemic spreading has a metacritical point defined by the awareness dynamics and the topology of multilayer networks. The structure of the interconnections between two networks significantly affects robustness~\cite{radicchi2013abrupt,
reis2014avoiding,van2016interconnectivity,cozzo2019layer}, synchronization \cite{aguirre2014synchronization,zhang2015explosive} and
spreading dynamics~\cite{hu2014conditions,de2017disease,
sanz2014dynamics}. Saumell-Mendiola \emph{et al.}~\cite{saumell2012epidemic}
demonstrated that interlayer degree correlations might trigger epidemic outbreaks. Wang
\emph{et al.}~\cite{wang2014asymmetrically} considered the coevolution of epidemics and
information spreading in multilayer networks, and it was demonstrated that
the interlayer degree correlations can also suppress
epidemic outbreaks without altering the outbreak threshold.

A natural question is to determine the optimal interlayer structure for spreading in multilayer networks. To address this, Aguirre \emph{et al.}~\cite{aguirre2013successful}
applied a matrix perturbation approach and
demonstrated that adding a connection between two nodes with large eigenvector centrality is
more likely to promote the spreading dynamics for
two competing networks. Recently, Pan \emph{et al.}~\cite{pan2019} suggested applying perturbation theory to the adjacency matrix to obtain the
optimal interconnections between two networks. This method is effective near the spreading threshold when a small number of edges are added.

In this study, we consider the problem of choosing a single interlayer edge that maximizes the spreading prevalence of the SIS model in two-layer networks. The SIS model can be applied to simple information or disease spreading. Therefore, understanding the maximization of the spreading prevalence may facilitate the promotion of information spreading or provide vaccination guidance~\cite{wang2016statistical}. We develop a theoretical framework that provides the optimal or near-optimal interconnecting edge for all parameter regions. Starting with the discrete Markovian chain approach for the SIS model in two isolated networks, we propose a perturbation method so that the spreading prevalence in the interconnected network may be accurately approximated. The edge with the largest incremental spreading prevalence is then chosen as the interconnecting edge. The incremental spreading prevalence incorporates information regarding both network structure and spreading dynamics. Moreover, it has a simple physical interpretation as a variant of Katz centrality~\cite{newman2010networks}, where the adjacency matrix is replaced by a matrix with weights depending on the dynamical information of the spreading process.

The paper is organized as follows. We present the
model in Sec.~\ref{sec:model} and then develop a theory
for obtaining the optimal interconnecting
strategy in Sec.~\ref{theory}. In Sec. \ref{num}, we perform extensive
numerical simulations to verify the effectiveness of the proposed
strategy. Sec.~\ref{con}
concludes the paper.

\section{Model description} \label{sec:model}
We consider the SIS model in two-layer networks. Let $a$ and $b$ be the two layers respectively. The number of nodes in $a$ and $b$ is denoted by $N_a$ and $N_b$, respectively, and the number of edges by $M_a$ and $M_b$, respectively. The adjacency matrices of the two layers are $G_a$ and $G_b$, and we assume that there are no interconnecting edges between them. Let $N={N_a+N_b}$. Then, the adjacency matrix $G^0$ of the two isolated layers combined is the following $N\times N$ matrix:
\begin{equation}
G^0=\left(
{\begin{array}{cc}
G_a & 0 \\
0 & G_b
\end{array}}
\right).
\end{equation}
We note that $0$ in the off-diagonal part denotes zero matrices. There are multiple ways to interconnect the two isolated networks, and the dynamics of the interconnected network rely on the interlayer structure. Our aim is to determine the optimal interconnecting edge such that the spreading prevalence is maximized.

By adding the interconnecting edge, the adjacency matrix becomes
\begin{equation}\label{add}
G=G^0+\delta G,
\end{equation}
where
\begin{equation}
\delta G=\left(
{\begin{array}{cc}
0 & G_{ab} \\
G_{ba} & 0
\end{array}}
\right)
\end{equation}
is the adjacency matrix for the interconnection between the two isolated networks. When $(G_{ab})_{ij}=(G_{ba})_{ji}=1$ for $i\in\{1,\cdots,N_a\}$ and $j\in\{1,\cdots,N_b\}$, an undirected edge is added between nodes $i$ and $j$.

We adopt the classical SIS model as the spreading model. Thus, each node can be in either the susceptible or infected state. Initially, a small fraction of nodes are selected as infected seeds, and the remaining nodes are susceptible. At each time step, every infected node in $a$
($b$) tries to infect the susceptible neighbors in the same network with probability $\lambda_a$
($\lambda_b$) and infect susceptible neighbors in
$b$ ($a$) with probability
$\lambda_{ab}$ ($\lambda_{ba}$). Then, all the infected nodes return to the susceptible state with probability $\gamma_{a}\ (\gamma_b)$. We assume
$\lambda_a=\lambda_b=
\lambda_{ab}=\lambda_{ba}=\lambda$ and $\gamma_a=\gamma_b$.
In the limit, the system reaches the steady state, and the fraction of infected nodes fluctuates around a stable value. Our aim is to choose an interconnecting edge such that the infected density of the new steady state in the interconnected network is maximized.

\section{Theoretical analysis} \label{theory}
To study the SIS model in networks, we adopt the discrete Markovian chain (DMC) approach~\cite{gomez2010discrete}, which
assumes that there are no dynamical correlations
among the states of neighbors~\cite{wang2016unification}. In this section, we first present the DMC approach for the SIS model in the network $G^0$ when there are no interconnections between the networks $a$
and $b$. Then, using a perturbation
method for DMC, we derive a formula that approximately provides the spreading prevalence in the interconnected network. Subsequently, we discuss physical interpretations of this formula, and finally, we study the problem of determining the optimal interconnecting edge based on the obtained formula.

\subsection{Perturbation method for the discrete Markovian chain}\label{subsecA}
Let $p_i(t)$ be the probability that node $i$ is infected at time $t$. Then, the node is susceptible with probability $1-p_i(t)$. If $i$ is in infected state at $t+1$, then  either it was infected at $t$ and has not recovered, or it was susceptible at $t$ and has been infected by at least one infected neighbor. The former case occurs with probability $(1-\gamma)p_i(t)$ and the latter with probability $(1-p_i(t))\left(1-q_i(t)\right)$. Here, $1-q_i(t)$ is the probability that node $i$ is infected by at least one infected neighbor at time $t$, which is given by
\begin{equation}\label{p_i}
q_i(t)=\prod_{j=1}^{N}[1-\lambda G_{ij}^0p_j(t)].
\end{equation}
Combining the two cases, the evolution equation of
$p_i(t)$ can be written as
\begin{equation}\label{i_t}
p_i(t+1)=(1-\gamma)p_i(t)+(1-p_i(t))\left(1-q_i(t)\right).
\end{equation}
In the steady state, we have $p_i(t)=p_i(t+1)=
p_i^*$ and $q_i(t)=q_i(t+1)=q_i^*$.
Writing Eqs.~(\ref{p_i}) and (\ref{i_t}) in terms of vectors in the steady state yields
\begin{equation}\label{eq:fixEq1}
p^*=(1-\gamma)p^*+(1-p^*)\circ (1-q^*)
\end{equation}
and
\begin{equation}\label{eq:fixEq2}
q^*_i=\prod_{j=1}^N(1-\lambda G_{ij}^0p^*_j),
\end{equation}
where $p^*$, $q^*$ are vectors of length $N$ with entries $p^*=(p_1^*,\cdots,p_N^*)^{\mathrm{T}}$,
$q^*=(q_1^*,\cdots,q_N^*)^{\mathrm{T}}$, and $\circ$ denotes component-wise vector product.
The expected number of infected nodes in the steady state is
\begin{equation}
\mathcal{P}= N^{-1} \mathbf{1}^{\mathrm{T}}p^*.
\end{equation}
Previous studies~\cite{gomez2010discrete,
de2017disease} demonstrated that a globally spreading outbreak occurs when the effective transmission probability $\lambda^*=\lambda/\gamma$ is larger than $1/\omega_1$, where $\omega_1$ is the leading eigenvalue of adjacency matrix $G^0$. That is, the spreading outbreak threshold is $\lambda^*_c=1/\omega_1$, whereas if $\lambda^*_c\leq 1/\omega_1$, then no outbreaks will be observed.

We now add one interconnecting edge between the two networks. Clearly, the spreading prevalence will increase after the edge is added. Subsequently, we develop a perturbation method to obtain an approximate estimate of the incremental spreading prevalence in the interconnected network.

When an interconnection is added between the two isolated networks, the adjacency matrix becomes $G=G^0+\delta G$. The fixed point of $p(t)$ in the interconnected network deviates from $p^*$ and the magnitude of deviation depends on where the interconnection is added. Nevertheless, as long as the two isolated networks are large enough, the modification in network structure can be regarded as small. As a consequence, the fixed point of $p(t)$ in the interconnected network should stay close to $p^*$. Since we focus on the case of adding one interconnection, this assumption should be valid for moderate network size. The actual magnitude of incremental spreading prevalence by interconnecting the networks can be seen from numerical results in Sec.~\ref{num}, for example, in Fig.~\ref{fig1}.

We now iterate the DMC equations in the interconnected network with initial condition $p(0)=p^*$, and then we use the decompositions $p(t)=p^*+\delta p (t)$ and $q(t)=q^*+\delta q(t)$ for some small $\delta p (t)$ and $\delta q (t)$. More explicitly, Eq.~(\ref{i_t}) in the interconnected network becomes
\begin{equation} \label{add_p}
\begin{split}
p^*+\delta p(t+1)&=(1-\gamma)(p^*+\delta p(t))\\
&+(1-p^*-\delta p(t))\circ (1-q^*-\delta q(t)).
\end{split}
\end{equation}
Expanding Eq.~(\ref{add_p}) and substituting Eq.~(\ref{eq:fixEq1}) yields
\begin{equation}\label{eq:fixPertuEq1}
\delta p(t+1)=(q^*-\gamma)\delta p(t)-(1-p^*)\cdot \delta q(t).
\end{equation}
We note that as $\delta p(t)$ and $\delta q(t)$ are assumed small, the second-order term $\delta p(t)\circ \delta q(t)$ is ignored. Similarly, Eq.~(\ref{p_i}) (the iteration equation for $q(t)$) in the interconnected network becomes
\begin{equation}\label{qi_f}
q_i^*+\delta q_i(t)=\prod_{j=1}^N\left(1-\lambda (G^0_{ij}+\delta G_{ij})(p_j^*+\delta p_j(t))\right).
\end{equation}
As before, by expanding this equation up to first-order terms in $\delta p(t)$, we obtain
\begin{equation}\label{eq:dqExpress}
\begin{split}
\delta q(t)=&-\lambda q^*\circ \left(G^0 +\delta G\right) Z \delta p(t)\\
&+q^*\circ \delta G\log(1-\lambda p^*),
\end{split}
\end{equation}
where $\log(1-\lambda p^*)$ is the vector obtained by taking the logarithm in each entry of $1-\lambda p^*$, and $Z$ is the $N\times N$ diagonal matrix  with entries
\begin{equation}\label{eq:Zdef}
Z_{ij} =\delta_{ij}\frac{1}{1-\lambda p_j^*}.
\end{equation}
The detailed derivation of Eq.~\eqref{eq:dqExpress} is provided in Appendix A.

Substituting Eq.~\eqref{eq:dqExpress} back into Eq.~\eqref{eq:fixPertuEq1} yields the following iteration formula for $\delta p(t)$:
\begin{equation}
\begin{split}
\delta p(t+1)=&(q^*-\gamma)\delta p(t) \\
&+(1-p^*)\circ\lambda q^*\circ \left(G^0 +\delta G\right) Z \delta p(t) \\
&-(1-p^*)\circ q^*\circ \delta G\log(1-\lambda p^*).
\end{split}
\end{equation}
This equation can be written in terms of matrix multiplication as follows:
\begin{equation}\label{eq:pertuRelationMatrix}
\delta p(t+1)=X\delta p(t)+y,
\end{equation}
where
\begin{equation}\label{A}
X=\lambda \mathrm{diag}(q^*-p^*\circ q^*)(G^0+\delta G)Z+\mathrm{diag}(q^*-\gamma)
\end{equation}
and
\begin{equation}\label{B}
y=-(1-p^*)\circ q^*\circ \delta G\log(1-\lambda p^*).
\end{equation}
Here, $\mathrm{diag}\left(\cdot\right)$ denotes the diagonal matrix with the elements of the input vector as diagonal entries. The stationary solution $\delta p^*$ of the perturbed system satisfies
\begin{equation}
\delta p^*=X\delta p^*+y,
\end{equation}
or in the closed form
\begin{equation}\label{eq:dprelation}
\delta p^*=\left(\mathbb{I}-X\right)^{-1}y.
\end{equation}
This provides an explicit relation between the interconnection edge and the stationary infected density increment. Therefore, it remains to choose $\delta G$ such that the incremental spreading prevalence
\begin{equation}\label{eq:dp}
\delta\mathcal{P}\mathop{:}=N^{-1}\mathbf{1}^{\mathrm{T}}\delta p=N^{-1}\mathbf{1}^{\mathrm{T}}\left(\mathbb{I}-X\right)^{-1}y
\end{equation}
is maximized. We note that Eq.~\eqref{eq:dp} holds even when we add multiple interconnecting edges.

\subsection{Physical interpretations} \label{int}
Before analytically studying the optimization of Eq.~\eqref{eq:dp}, we should intuitively understand which interconnecting edge will give larger incremental spreading prevalence. We recall the Katz centrality~\cite{newman2010networks} $S_{\mathrm{Katz}}$, which is defined by
\begin{equation}
S_{\mathrm{Katz}}=\mathbf{1}^{\mathrm{T}}\left(\mathbb{I}-\beta G\right)^{-1},
\end{equation}
where $G$ is the adjacency matrix and $\beta$ a tunable parameter. Then, $S_{\mathrm{Katz}}$ is a vector with entries representing the centrality of the corresponding nodes.
Katz centrality is defined by considering the number of weighted walks between nodes, where $\beta$ is the attenuation factor of walk length~\cite{newman2010networks}. The matrix inverse in $S_{\mathrm{Katz}}$ has expansion
\begin{equation}\label{eq:katzExpansion}
\left(\mathbb{I}-\beta G\right)^{-1}=\mathbb{I}+\beta G+\beta^2 G^2+\cdots,
\end{equation}
where $G^{t}$ is the matrix multiplication of $G$ by itself $t$ times. The entry $G^t_{ij}$ of $G^t$ then counts the number of walks of length $t$ between nodes $i$ and $j$.

We define a row vector
\begin{equation}
S_{\mathrm{Dyn}}=\mathbf{1}^{\mathrm{T}}\left(\mathbb{I}-X\right)^{-1}.
\end{equation}
Then, Eq.~\eqref{eq:dp} can be written as $\delta\mathcal{P}=N^{-1} S_{\mathrm{Dyn}} y$. The vector $S_{\mathrm{Dyn}}$ has the same form as $S_{\mathrm{Katz}}$, with $\beta G$ in $S_{\mathrm{Katz}}$ replaced by $X$. We now further explore this connection and interpret $S_{\mathrm{Dyn}}$ as a weighted version of $S_{\mathrm{Katz}}$.

By the definition in Eq.~\eqref{A}, the entries of $X$ are given by
\begin{equation}\label{eq:Xij}
X_{ij}=G_{ij} \frac{\lambda \left(1-p^*_i\right)q^*_i}{1-\lambda p^*_j}
\end{equation}
for $i\neq j$. We note that $X_{ij}$ is nonzero only when $G_{ij}$ is nonzero; thus, $X$ can be understood as a weighted network with edge weights defined in terms of the dynamical information provided by $p^*$ and $q^*$. By Eq.~\eqref{eq:Xij}, it is straightforward that the edge weight $X_{ij}$ is a decreasing function of $p^*_i$ and an increasing function of $q^*_i,p^*_j$. That is, the edge connecting $i$ and $j$ will have a larger weight if it is less infected at $i$ (small $q^*_i$) and a neighborhood of $i$ (large $q^*_i$) but highly infected at $j$ (large $p^*_j$). More briefly, edges connecting less infected and highly infected regions will have larger weights.

Accordingly, $S_{\mathrm{Dyn}}$ is simply the Katz index defined on the weighted graph. $S_{\mathrm{Katz}}$ does not distinguish walks with the same length, as can be seen from Eq.~\eqref{eq:katzExpansion}, whereas $S_{\mathrm{Dyn}}$ further weights walks using dynamical information. When the transmission probability is below the spreading threshold, we have $p^*\approx 0$ and $q^*\approx 1$. Then, $X\approx 1-\gamma+\lambda G$ and
\begin{equation}
\left(\mathbb{I}-X\right)^{-1}\approx \gamma^{-1} \left(\mathbb{I}-
\frac{\lambda}{\gamma}G\right)^{-1}.
\end{equation}
In this case, $S_{\mathrm{Dyn}}$ reduces to $S_{\mathrm{Katz}}$ with $\beta=\lambda/\gamma$ (up to a constant factor $\gamma^{-1}$).

The incremental spreading prevalence $\delta\mathcal{P}=N^{-1} S_{\mathrm{Dyn}} y$ is then the weighted average of $S_{\mathrm{Dyn}}$, with nodes again weighted by the vector $y$. By the definition in Eq.~\eqref{B}, the entries of $y$ are
\begin{equation}
y_i=-(1-p^*_i)q^*_i\sum_{j=1}^N\delta G_{ij} \log \left(1-\lambda p^*_j\right).
\end{equation}
Similarly, $y_i$ is a decreasing function of $p^*_i$ and an increasing function of $q^*_i$ and $p^*_j$. Thus, $y_i$ takes lager values if $i$ and its neighborhood are less infected and is connected to a highly infected node by an interconnection in $\delta G$.

By combining the discussions on $X$ and $y$, the optimal strategy can be understood as selecting an edge such that the infection is more easily transmitted from highly infected to less infected regions, which is consistent with intuition. We will refer to the method proposed in this section as \textit{dynamical Katz method}.

\subsection{Choosing the optimal edge}\label{sec:optimize}
We now discuss the optimization of Eq.~\eqref{eq:dp}. We will consider only one connecting edge between the two isolated networks, that is, the optimal edge. We first introduce some notations. For the vector $p^*$, let $p^*_a$ be its part corresponding to network $a$. Specifically, $p^*_a$ is a vector of length $N_a$ with elements
\begin{equation}
\left(p^*_a\right)_i=\left(p^*\right)_i
\end{equation}
for $1\leq i\leq N_a$. $p^*_b$, $q^*_a$, $q^*_b$, $y_a$, and $y_b$ are defined analogously. We define the $N_a\times N_a$ diagonal matrix $Z_a$ with entries
\begin{equation}
(Z_a)_{ik} =\delta_{ik}\frac{1}{1-\lambda \left(p^*_a\right)_i}=Z_{ik},
\end{equation}
for $1\leq i,k \leq N_a$. $Z_b$ is defined analogously and corresponds to $b$.

We decompose $X$ as $X=X^0+\delta X$, where
\begin{equation}
X^0 =\lambda \mathrm{diag}(q^*-p^*\circ q^*)G^0 Z+\mathrm{diag}(q^*-\gamma)
\end{equation}
depends only on $G^0$, and
\begin{equation}
\delta X =\lambda \mathrm{diag}(q^*-p^*\circ q^*)\delta G Z
\end{equation}
depends only on $\delta G$. We note that $X^0$ is a diagonal block matrix and can be further written as
\begin{equation}
X^0=\left(
{\begin{array}{cc}
X^0_a & 0 \\
0 & X^0_b
\end{array}}
\right),
\end{equation}
where $X^0_a$ is the block diagonal part of $X^0$ that depends only on $G_a$, with
\begin{equation}
X^0_a=\lambda \mathrm{diag}(q^*_a-p^*_a\circ q^*_a)G_a Z_a+\mathrm{diag}(q^*_a-\gamma).
\end{equation}
Similarly,
\begin{equation}
X^0_b=\lambda \mathrm{diag}(q^*_b-p^*_b\circ q^*_b)G_b Z_b+\mathrm{diag}(q^*_b-\gamma).
\end{equation}
$\delta X$ is an off-diagonal block matrix
\begin{equation}
\delta X=\left(
{\begin{array}{cc}
0 & \delta X_{ab} \\
\delta X_{ba} & 0
\end{array}}
\right),
\end{equation}
with the off-diagonal blocks given by
\begin{equation}
\begin{split}
\delta X_{ab}&=\lambda \mathrm{diag}(q^*_a-p^*_a\circ q^*_a)G_{ab} Z_b\\
\delta X_{ba}&=\lambda \mathrm{diag}(q^*_b-p^*_b\circ q^*_b)G_{ba} Z_a.
\end{split}
\end{equation}
Using the properties of block matrices, the matrix inverse in Eq.~\eqref{eq:dprelation} can be written as
\begin{equation}
\begin{split}
\left(\mathbb{I}-X\right)^{-1}=&\left(
{\begin{array}{cc}
\mathbb{I}-X^0_a & -\delta X_{ab} \\
-\delta X_{ba} & \mathbb{I}-X^0_b
\end{array}}
\right)^{-1}\\
=&\left(
{\begin{array}{cc}
C & C\delta X_{ab}B \\
D \delta X_{ba}A & D
\end{array}}
\right),
\end{split}
\end{equation}
where
\begin{equation}
A=\left(\mathbb{I}-X^0_a\right)^{-1},\ B=\left(\mathbb{I}-X^0_b\right)^{-1}
\end{equation}
and
\begin{equation}
\begin{split}
C&=\left(\mathbb{I}-X^0_a-\delta X_{ab} B \delta X_{ba}\right)^{-1},\\
D&=\left(\mathbb{I}-X^0_b-\delta X_{ba} A \delta X_{ab}\right)^{-1}.
\end{split}
\end{equation}

We now assume that we add an interconnecting edge between node $i$ of network $a$ and node $j$ of $b$. By the Sherman--Morrison formula, the resulting increment in the spreading prevalence can be written in the following explicit form:
\begin{equation} \label{strag}
\begin{split}
N\delta \mathcal{P}=&\frac{c_{ij}+c_{ji} x_{ij} B_{jj}}{1-x_{ij}x_{ji}A_{ii}B_{jj}} \left(\mathbf{1}^{\mathrm{T}}A\right)_i
\\
&+\frac{c_{ji}+c_{ij} x_{ji} A_{ii}}{1-x_{ij}x_{ji}A_{ii}B_{jj}} \left(\mathbf{1}^{\mathrm{T}}B\right)_j,
\end{split}
\end{equation}
where
\begin{equation}\label{eq:xDef}
\begin{split}
x_{ij}&\mathop{:}=\lambda \left[\left(q^*_a\right)_i-\left(p^*_a\right)_i \left(q^*_a\right)_i\right]\left[1-\lambda \left(p^*_b\right)_j\right]^{-1},\\
x_{ji}&\mathop{:}=\lambda \left[\left(q^*_b\right)_j-\left(p^*_b\right)_j \left(q^*_b\right)_j\right]\left[1-\lambda \left(p^*_a\right)_i\right]^{-1}
\end{split}
\end{equation}
and
\begin{equation}\label{eq:cDef}
\begin{split}
c_{ij}&=-\left[\left(q^*_a\right)_i-\left(p^*_a\right)_i \left(q^*_a\right)_i\right]\log\left[1-\lambda \left(p^*_b\right)_j\right],\\
c_{ji}&=-\left[\left(q^*_b\right)_j-\left(p^*_b\right)_j \left(q^*_b\right)_j\right]\log \left[1-\lambda \left(p^*_a\right)_i\right].
\end{split}
\end{equation}
The detailed derivation of Eq.~\eqref{strag} is given in Appendix B.

This provides a simple formula for the spreading prevalence in the interconnected network. The optimal strategy is simply to select the edge with the highest corresponding $\delta\mathcal{P}$.
This strategy relies not only on the network topology (i.e., the adjacency matrices $G_a$ and $G_b$) but also on the dynamical information of the spreading process when the two networks are isolated (i.e., $\lambda$, $\gamma$, $q^*$, and $p^*$).

\section{Numerical simulations} \label{num}
In this section, we perform extensive
numerical simulations on both synthetic and real-world networks to verify the performance of the strategy. We note
that we do not compare the DMC predictions with Monte Carlo simulations because the DMC approach can accurately predict the simulations~\cite{gomez2010discrete}. In the following, the numerical value of $\delta \mathcal{P}$ obtained by iterating the DMC is denoted by $\delta \mathcal{P}^{\mathrm{num}}$. The optimal edge predicted by the DMC equations is called the numerical optimal edge. The approximate $\delta \mathcal{P}$ predicted by Eq.~\eqref{strag} is denoted by $\delta \mathcal{P}^{\mathrm{approx}}$.

For two networks with number of nodes $N_a$ and $N_b$, there are in total $M_l=N_a\times N_b$ latent interconnections.
For small networks, it is possible to check all the latent connections exhaustively so that the optimal may be determined. However, for large $N$, an exhaustive search is slow and gradually becomes impossible. We first use small networks to verify the accuracy of $\delta \mathcal{P}^{\mathrm{approx}}$ predicted by Eq.~\eqref{strag} and compare the optimal edge by this strategy with the numerical optimal edge (the optimal edge predicted by the DMC equations).

To construct synthetic networks, we adopt the uncorrelated configuration model with power-law degree distributions. Specifically, we set the degree distributions of networks $a$ and $b$ to $P(k)\sim k^{-\alpha_a}$
and $P(k)\sim k^{-\alpha_b}$ respectively,
where $\alpha_a$ and $\alpha_b$ are the degree exponents. The network sizes are set to $N_a=N_b=100$. Without loss of generality, we set the recovery probability of the SIS model to $\gamma=0.5$ and make the infection probability $\lambda$ a tunable parameter.

\begin{figure}
\begin{center}
\epsfig{file=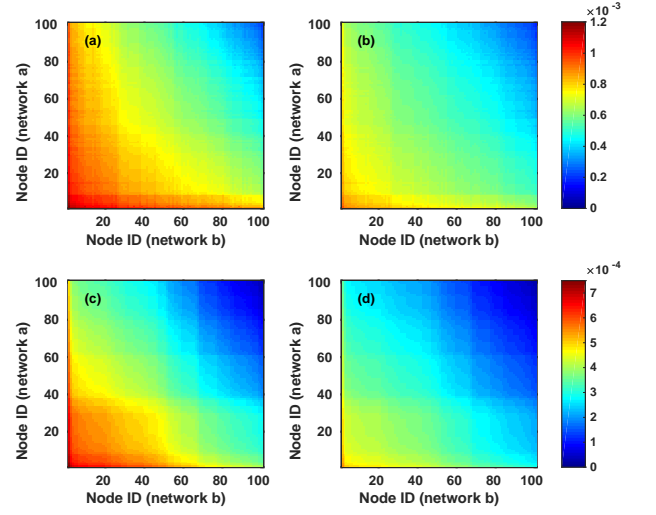,width=1\linewidth}
\caption{(Color online) Incremental spreading prevalence $\delta \mathcal{P}$ by adding each latent interconnection separately. The vertical and horizontal axes correspond to node IDs in the networks. Thus, each point in the plots corresponds to an edge connecting $a$ and $b$, and its color represents the value of $\delta \mathcal{P}$ by adding the edge. (a)  Approximate and (b) numerical predictions of $\delta \mathcal{P}$ with $\lambda=0.3$.
(c) Approximate and (d) numerical predictions of $\delta \mathcal{P}$ with $\lambda=0.5$. The nodes are arranged in identical order for (a) and (b), as well as for (c) and (d). Other parameters are set as $N_a=N_b=100$, $\alpha_a=2.3$, $\alpha_b=3.0$ and $\gamma=0.5$.}
\label{fig1}
\end{center}
\end{figure}

We first compare $\delta \mathcal{P}^{\mathrm{approx}}$ predicted by Eq.~\eqref{strag} with the predictions by the DMC approach. For each latent edge connecting node $i\in \{1,\cdots N_a\}$ and node $j\in\{1,\cdots,N_b\}$, we compute $\delta \mathcal{P}^{\mathrm{approx}}$ using Eq.~\eqref{strag} for $\lambda=0.3$ (Fig.~\ref{fig1}(a)) and $\lambda=0.5$ (Fig.~\ref{fig1}(c)). Then, we add the edge to the network and iterate the DMC to obtain $\delta \mathcal{P}^{\mathrm{num}}$, which is shown in Figs.~\ref{fig1}(b) and ~\ref{fig1}(d) for $\lambda=0.3$ and $\lambda=0.5$, respectively. Nodes are arranged in identical order in Figs.~\ref{fig1}(a), \ref{fig1}(b), also in identical order in Figs.~\ref{fig1}(c),~\ref{fig1}(d). The approximate values usually appear higher than the numerical values, but intuitively, they are strongly correlated. The maximum relative error $\left(\delta \mathcal{P}^{\mathrm{approx}}-\delta \mathcal{P}^{\mathrm{num}}\right)/\delta \mathcal{P}^{\mathrm{num}}$ for all edges is $0.315$ in Figs.~\ref{fig1}(a) and (b), and $0.396$ in Figs.~\ref{fig1}(c) and (d). However, we will demonstrate that they are almost linearly correlated in order, which suggests the approximate value is sufficient to obtain the optimal edge.

\begin{figure}
\begin{center}
\epsfig{file=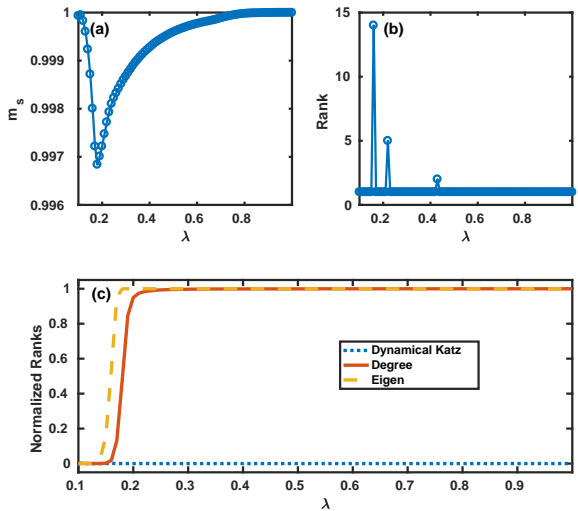,width=0.98\linewidth}
\caption{(Color online) Performance of different
strategies versus transmission probability. (a) Spearman's rank correlation coefficient $m_s$ between ranks predicted by the dynamical Katz method and the numerical ranks. (b) Numerical rank of the optimal edge predicted by the dynamical Katz method. (c) Normalized numerical rank of the optimal edge predicted by dynamical Katz (blue dotted line), degree (orange solid line), and eigenvector centrality (yellow dashed line) methods. Below the spreading threshold, the prevalence $\mathcal{P}$ is zero, and the rankings are trivial; thus, we consider $\lambda$ only in the range starting slightly above the threshold. Other parameters are set as
$N_A=N_B=100$, $\alpha_a=3.0$, $\alpha_b=2.3$ and $\gamma=0.5$.}
\label{fig2}
\end{center}
\end{figure}

To see the correlations, we compute the Spearman's rank correlation coefficient~\cite{lee2012correlated,wang2014asymmetrically} between the approximate and numerical $\delta \mathcal{P}$. We score all the latent interconnecting edges by their $\delta \mathcal{P}^{\mathrm{num}}$ and $\delta \mathcal{P}^{\mathrm{approx}}$; then, two rankings can be obtained. Let $r_{ij}$ and $r^{\prime}_{ij}$ be the rank of the edge connecting node $i$ in network $a$ and node $j$ in network $b$ scored by $\delta \mathcal{P}^{\mathrm{num}}$ and $\delta \mathcal{P}^{\mathrm{approx}}$, respectively.  Spearman's rank correlation coefficient is defined as
\begin{equation}\label{ms}
m_s=1-6\frac{\sum_{i=1}^{N_a} \sum_{j=1}^{N_b} \left(r_{ij}-r^{\prime}_{ij}\right)^2}{M_l(M_l^2-1)}.
\end{equation}
We plot $m_s$ as a function of $\lambda$ in Fig.~\ref{fig2}(a). It can be observed that Spearman's rank correlation coefficients are close to $1$ for all $\lambda$. The minimum value of $m_s$ for all $\lambda$ in Fig.~\ref{fig2}(a) is $0.9968$. This suggests that the proposed strategy accurately predicts the overall order of $\delta P^{\mathrm{num}}$.

In addition to the strong overall correlations for the approximate and numerical values of $\delta \mathcal{P}$, we are particularly concerned with the top-ranked edge. We further verify the performance of the strategy by comparing the predicted optimal edge with the numerical optimal edge. For each $\lambda$, we select the edge with the highest $\delta \mathcal{P}^{\mathrm{approx}}$ predicted by the dynamical Katz method and compute its numerical rank in all the latent edges. The edge rank versus $\lambda$ is shown in Fig.~\ref{fig2}(b). It can be seen that the rank is $1$ or near $1$ for all values of $\lambda$. When the rank is exactly $1$, the optimal edge predicted by the dynamical Katz method coincides with the numerical optimal edge, and this is the case for most values of $\lambda$.

As the dynamical Katz strategy incorporates information regarding both the network structure and spreading dynamics, it is useful to compare it with simple strategies that consider only the static network structure to understand the role of dynamical information. Specifically, we consider the strategy of connecting the two nodes with the highest degree or eigenvector centrality. The normalized ranks (ranks divided by $M_l=N_a\times N_b$) by the dynamical Katz and the two static strategies are shown in Fig.~\ref{fig2}(c). All three strategies are optimal or near-optimal when the transmission probability $\lambda$ is slightly above the critical value, but the two static strategies fail quickly when $\lambda$ becomes large, whereas the dynamical Katz method still performs well.

As discussed in Sec.~\ref{int}, when $p^*\approx 0$, the dynamical Katz matrix reduces to the Katz matrix. When $\lambda/\gamma$ is small, we have
\begin{equation}\label{eq:alable}
\left(\mathbb{I}-X\right)^{-1}\approx \gamma^{-1}\mathbb{I}+\lambda G,
\end{equation}
and this reduces to degree centrality. For uncorrelated configuration models, degree and eigenvector centrality are strongly correlated. When $\lambda$ is small, nodes with high centrality values (i.e., degree and eigenvector centrality) have a larger probability to be infected. If we connect  them with an edge, then high-centrality nodes together with their neighbors could form an infected cluster~\cite{goltsev2012localization}
and further transmit the infection to other nodes.
Thus, for small $\lambda$, the degree and eigenvector strategies perform well. For large values of $\lambda$,
globally spreading outbreaks occur, and nodes with small
centrality have a higher probability to
be susceptible. In this case, additional
connections to these nodes are required for promoting the spreading
dynamics. Therefore, both the degree and eigenvector strategies
fail, and the dynamical information should be considered.

\begin{figure}
\begin{center}
\epsfig{file=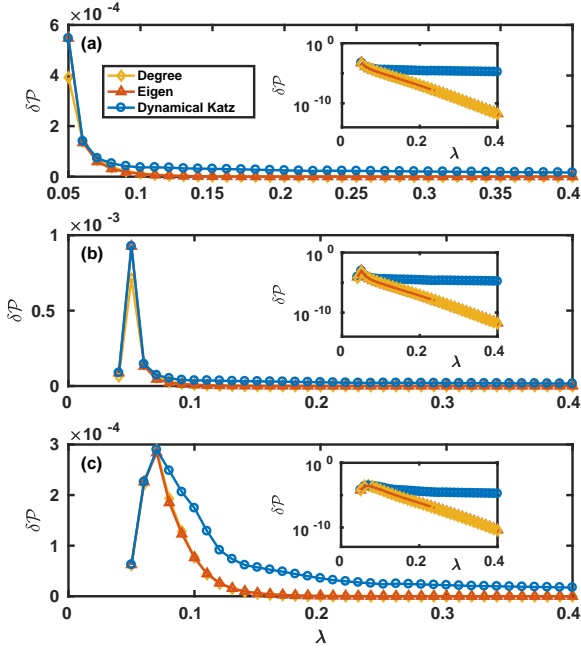,width=1\linewidth}
\caption{(Color online) Incremental spreading prevalence $\delta \mathcal{P}$ versus $\lambda$ in synthetic networks when one interconnecting edge is added. Networks with power-law degree distributions are considered, where the degree exponents are (a) $\alpha_a=2.3$, $\alpha_b=3.0$, (b) $\alpha_a=3.0$, $\alpha_b=3.0$, and (c) $\alpha_a=4.0$, $\alpha_b=3.0$. $\delta \mathcal{P}$ versus $\lambda$ in semi-log plots are shown in the insets, where $\delta \mathcal{P}$ is on the logarithm scale. Data points corresponding to the degree and eigenvector centrality strategies are highly overlapped. We consider $\lambda$ in the range starting slightly above the spreading threshold to avoid trivial cases. Other parameters are set as $N_A=N_B=5000$ and $\gamma=0.5$.}
\label{fig3}
\end{center}
\end{figure}

For large networks, exhaustive searching becomes impossible. In this case, we compare the performance of the dynamical Katz method with that of the two static methods based on degree and eigenvector centrality. For the three methods, we add the predicted optimal edge separately and compare the resulting $\delta \mathcal{P}$. We first consider synthetic networks. We construct three pairs of networks with power-law degree distributions, with degree exponents (\romannumeral 1) $\alpha_a=2.3$, $\alpha_b=3.0$, (\romannumeral 2) $\alpha_a=3.0$, $\alpha_b=3.0$, and (\romannumeral 3) $\alpha_a=4.0$, $\alpha_b=3.0$. The graphs of $\delta \mathcal{P}$ versus $\lambda$ for the three network pairs are shown in Fig.~\ref{fig3}. We further add the semi-log plot in the insets for the first two network pairs for better visualization, with $\delta \mathcal{P}$ on a logarithm scale.  When $\lambda$ is close to the critical point, all three strategies exhibit highly similar performance. As in small networks, this could also be near the maximal possible value of $\delta \mathcal{P}$. When $\lambda$ becomes large, dynamical Katz outperforms the other two static methods for all network pairs. In this case, connecting nodes with large degree or eigenvector centrality yields almost zero $\delta \mathcal{P}$, which decays with $\lambda$, as can be seen in, e.g., the insets in Figs.~\ref{fig3}(a) and (b). Moreover, it is worth noticing that $\delta \mathcal{P}$ is always maximized slightly above the spreading threshold, which suggests that the marginal improvement is maximized near the critical point.

We now test the dynamical Katz method on real-world networks. Three pairs of networks are considered: (\romannumeral 1) Advogato~\cite{Massa2009}, Facebook~\cite{Leskovec2012}, (\romannumeral 2) OpenFlights~\cite{Konect}, Air traffic control~\cite{Konect}, and (\romannumeral 3) Adolescent health~\cite{moody2001peer}, Physicians~\cite{coleman1957}.
The first pair (Advogato and Facebook) are two online social networks, the second pair (OpenFlights and Air traffic control) are infrastructure networks of airports and flights, and the third pair (Adolescent health and Physicians) are two offline social networks. The networks were downloaded from~\cite{Konect}, and details can be found therein. Some basic statistics are shown in TABLE~\ref{table1}.

\begin{figure}
\begin{center}
\epsfig{file=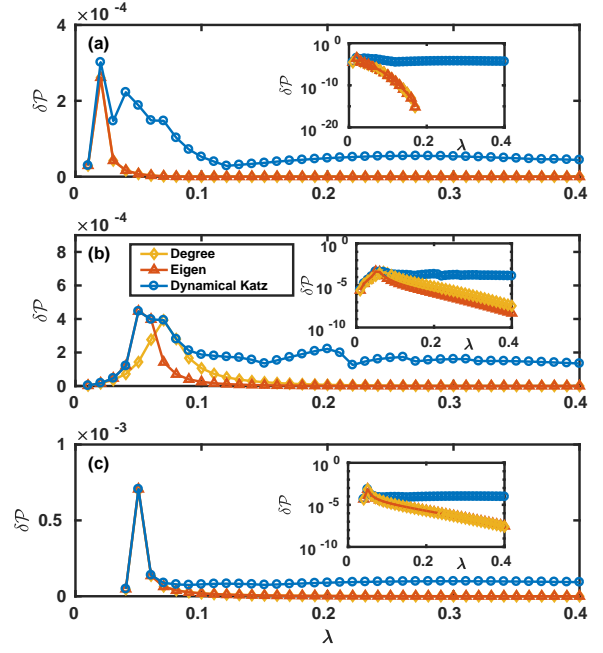,width=1\linewidth}
\caption{(Color online) Incremental spreading prevalence $\delta \mathcal{P}$ versus $\lambda$ when one interconnecting edge is added in real-world networks. The network pairs are (a) Advogato, Facebook~, (b) OpenFlights, Air traffic control, and (c) Adolescent health, Physicians. Data points for the degree and the eigenvector centrality strategies are highly overlapped for some values of $\lambda$. $\delta \mathcal{P}$ versus $\lambda$ in semi-log plots are shown in the insets, where $\delta \mathcal{P}$ is on the logarithm scale. In (a), for large values of $\lambda$, $\delta \mathcal{P}$ given by Degree and Eigen are so close to zero such that numerically we have $\delta \mathcal{P}=0$ due to the limitation of numerical accuracy. Therefore in the inset of (a), data points for large $\lambda$ cannot be seen on a logarithm scale.  Basic statistics of the six real-world networks can be found in TABLE~\ref{table1}.}
\label{fig4}
\end{center}
\end{figure}

$\delta \mathcal{P}$ versus $\lambda$ for the three network pairs are shown in Fig.~\ref{fig4}. As in the case of the synthetic networks, it can be seen that the dynamical Katz method performs best for all values of $\lambda$. However, for small values of $\lambda$, the three methods are quite close. For larger $\lambda$, the two static method yield $\delta \mathcal{P}$ very close to zero, whereas the dynamical Katz exhibits significant improvement. The results further confirmed the effectiveness of the dynamical Katz method on real-world networks.

\begin{table}[htbp]
\caption{Basic statistics of six real-world networks: number of nodes ($N$), number of edges ($M$), maximal degree ($k_{\mathrm{max}}$), first ($\langle k \rangle$) and second ($\langle k^2 \rangle$) moments of the degree distribution, and the theoretical spreading threshold predicted by DMC $\lambda_c^*=1/\omega_1$.}
\begin{tabular}{ccccccc}
\toprule
\hline
\hline
\ Networks & $N$ & $M$ & $k_{\mathrm{max}}$ & $\langle k \rangle$ & $\langle k^2 \rangle$ & $\lambda^*_c$ \\
\midrule
\hline
Advogato & 5042 & 39227 &  803 & 15.56 & 1284.00 & 0.014\\
Facebook & 2888 & 2981 & 769 &  2.06 &  528.13 & 0.036\\
OpenFlights & 2905 &  15645 & 242 &  10.77 & 601.45 & 0.016\\
Air traffic control & 1226 &  2408 & 34 & 3.928 & 28.90 & 0.109\\
Adolescent health & 2539 &  10455 & 27 & 8.24 & 86.41 & 0.076\\
Physicians & 117 &  465 & 26 & 7.95 & 79.16 & 0.099\\
\bottomrule
\hline
\hline
\end{tabular}
\label{table1}
\end{table}

\begin{figure}
\begin{center}
\epsfig{file=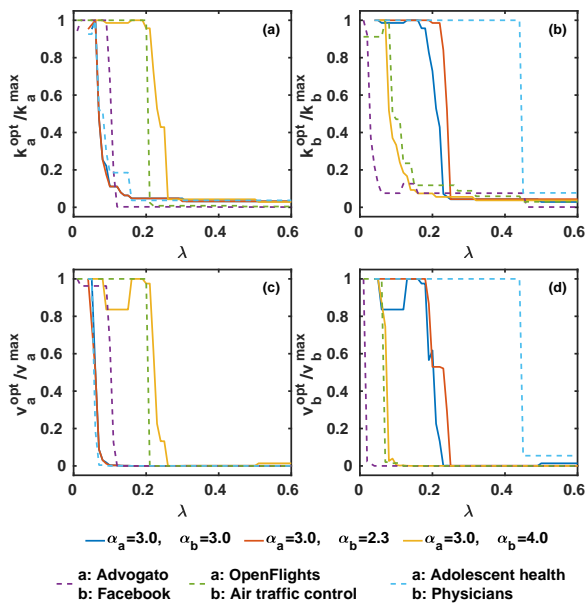,width=1\linewidth}
\caption{(Color online) Normalized degrees and eigenvector centralities of the optimal edge's two end-nodes versus $\lambda$ for both synthetic and real-world networks. (a) $k_a^{\mathrm{opt}}/k_a^{\mathrm{max}}$, i.e., normalized degree of the optimal edge's end-node in layer $a$, versus $\lambda$. (b) $k_b^{\mathrm{opt}}/k_b^{\mathrm{max}}$, i.e., normalized degree of the optimal edge's end-node in layer $b$, versus $\lambda$. (c) $v_a^{\mathrm{opt}}/v_a^{\mathrm{max}}$, i.e., normalized eigenvector centrality of the optimal edge's end-node in layer $a$, versus $\lambda$. (d) $v_b^{\mathrm{opt}}/v_b^{\mathrm{max}}$, i.e., normalized eigenvector centrality of the optimal edge's end-node in layer $b$, versus $\lambda$. Different lines correspond to different network pairs, among which the solid lines correspond to synthetic networks and dashed lines correspond to real-world networks. The values of $\lambda$ considered start slightly above the spreading threshold of each pair of networks.}
\label{fig5}
\end{center}
\end{figure}

For both synthetic and real-world networks, the Degree and Eigen methods work well near the critical $\lambda$ and fail when $\lambda$ becomes large, while the dynamical Katz method performs well in all the parameter region. To better understand the structural properties of the optimal edge predicted by the dynamical Katz method, we study how the degrees and eigenvector centralities of the optimal edge's two end-nodes change with $\lambda$. Let $k_a^{\mathrm{opt}}$ and $k_b^{\mathrm{opt}}$ be the degrees of the optimal edge's two end-nodes in layer $a$ and layer $b$ respectively. Similarly, we define $v_a^{\mathrm{opt}}$ and $v_b^{\mathrm{opt}}$ as the two nodes' eigenvector centralities. Let $k_a^{\mathrm{max}}$ and $k_b^{\mathrm{max}}$ be the maximum degree of $a$ and $b$ respectively, while  $v_a^{\mathrm{max}}$ and $v_b^{\mathrm{max}}$ be the maximum eigenvector centrality of $a$ and $b$ respectively.
$k_a^{\mathrm{opt}}/k_a^{\mathrm{max}}$, $k_b^{\mathrm{opt}}/k_b^{\mathrm{max}}$, $v_a^{\mathrm{opt}}/v_a^{\mathrm{max}}$ and $v_b^{\mathrm{opt}}/v_b^{\mathrm{max}}$ versus $\lambda$ are shown in Figs.~\ref{fig5}(a)-(d) respectively. When $\lambda$ is near the critical point, nodes with high degree and eigenvector centrality are chosen to be connected. Near the critical point, the spreading prevalence is small, connecting nodes with high centrality will help to maintain the infected cluster and further transmit the infection to other nodes. When $\lambda$ becomes large, nodes with high centrality have a very high probability to be infected, therefore connecting these nodes becomes unnecessary. As shown in Fig.~\ref{fig5}, the degrees and eigenvector centralities of the optimal edge become small when $\lambda$ becomes large. The numerical results have further verified the discussions about the relations between dynamical Katz and other two static methods (below Eq.~\eqref{eq:alable}).

\section{Discussion} \label{con}
We studied the problem of determining the optimal interconnecting edge for promoting spreading dynamics. By applying a perturbation method to the DMC equations, we obtained a Katz-like index for predicting the spreading prevalence in the interconnected networks. This index accurately predicts the optimal interconnecting edge for promoting spreading over all parameter regions, as demonstrated in small networks. For large synthetic and real-world networks, the method outperformed certain static strategies, namely, connecting nodes with highest degree or eigenvector centrality. For small $\lambda$, the three strategies had similar performance. For large $\lambda$, the two heuristic strategies yielded almost zero incremental spreading prevalence, whereas the dynamical Katz method performed well. In addition to accurately predicting the optimal edge, the dynamical Katz method provides a clear physical interpretation of how the optimal edge is chosen.

We considered the addition of only one interconnecting edge, but real-world multilayer networks usually have multiple interconnecting edges. We note that Eq.~\eqref{eq:dp}, which estimates the incremental spreading prevalence in terms of interconnections, is valid for general interconnecting structures. This could provide the foundation for further study in the case of multiple edges. For the single-edge case, the interconnection matrix $C$ can be written as the outer product of two vectors. By applying the Sherman--Morrison formula, $\delta \mathcal{P}$ can take a simple form that is easy to optimize. When multiple edges are added, the outer product decomposition of $C$ cannot be used in general.

A simple heuristic method for adding multiple edges is to add edges one by one using the proposed method. Specifically, at each step, one edge is added using the dynamical Katz method, and then the DMC equations are iterated to converge in this new network. The procedure is repeated until all edges are added. By adding one edge, the dynamical Katz method is likely to be optimal or near-optimal; therefore, this heuristic algorithm can be considered a greedy algorithm. The performance of such a greedy algorithm can be further analyzed, and more sophisticated algorithms could be designed. We leave this as an open issue for future exploration. Moreover, the perturbation method developed in this study could also be extended to other types of networks (e.g., temporal networks) and spreading models (e.g., social contagions and cascading failures).

\begin{acknowledgments}
This work was partially supported by National Natural Science Foundation of China (Nos.~61433014 and 61673086), China Postdoctoral Science Foundation (Grant No.~2018M631073), China Postdoctoral Science Special Foundation (Grant No.~2019T120829) and Fundamental Research Funds for the Central Universities.
\end{acknowledgments}

\section*{Appendix A: Derivation of the perturbuted equation for $\delta q(t)$}\label{sec:appendixa}
\setcounter{equation}{0}
\renewcommand\theequation{A.\arabic{equation}}
In this section we detail the derivation of the perturbed equation for $\delta q(t)$ in Eq.~\eqref{eq:dqExpress} starting from Eq.~(\ref{qi_f}). Since $G^0$ is a diagonal block matrix, while $\delta G$ is an off-diagonal block matrix, then $G^0_{ij}=1$ and $\delta G_{ij}=1$ can not be observed simultaneously. Thus the following equation holds
\begin{equation}\nonumber
\begin{split}
&\left(1-\lambda (G^0_{ij}+\delta G_{ij})\right)\left(p_j^*+\delta p_j(t)\right)\\
=&\left(1-\lambda G^0_{ij}(p_j^*+\delta p_j(t))\right)\left(1-\lambda \delta G_{ij}(p_j^*+\delta p_j(t))\right),
\end{split}
\end{equation}
which can be checked by substituting all possible combinations of $G^0_{ij}$ and $\delta G_{ij}$.
Divide by $q^*_i$ for both sides and substitute Eq.~(\ref{eq:fixEq2}) gives
\begin{equation}\label{eq:pertuQ}
\begin{split}
1+\frac{\delta q_i(t)}{q_i^*}&=\prod_{j=1}^N\left(1-\frac{\lambda G^0_{ij}\delta p_j(t)}{1-\lambda G_{ij}^0p_{j}^*}\right)  \\
&\times \prod_{j=1}^N\left(1-
\frac{\lambda\delta G_{ij}\delta p_j(t)}{1-\lambda\delta G_{ij}p^*_{j}}\right)\prod_{j=1}^N\left(1-\lambda\delta G_{ij}p^*_{j}\right).
\end{split}
\end{equation}
Note that the following relation holds
\begin{equation}
\frac{\lambda G^0_{ij}\delta p_j(t)}{1-\lambda G_{ij}^0p_{j}^*}=G^0_{ij}\frac{\lambda \delta p_j(t)}{1-\lambda p_{j}^*},
\end{equation}
since $G^0_{ij}\in \{0,1\}$ and similarly when replacing $G^0_{ij}$ by $\delta G_{ij}\in \{0,1\}$. Take the logarithm on both sides of Eq.~(\ref{eq:pertuQ}), expand to the first orders of $\delta p_i(t)$, $\delta q_i(t)$, and apply the above relation gives
\begin{equation}\label{eq:pertuQ2}
\begin{split}
\frac{\delta q_i(t)}{q_i^*}=&-\sum_{j=1}^N G^0_{ij}\frac{\lambda \delta p_j(t)}{1-\lambda p_{j}^*}-\sum_{j=1}^N \delta G_{ij}\frac{\lambda \delta p_j(t)}{1-\lambda p_{j}^*}\\
&+\sum_{j=1}^N \log \left(1-\lambda\delta G_{ij}p^*_{j}\right).
\end{split}
\end{equation}
Again the terms in the last summation can be checked satisfying
\begin{equation}
\log \left(1-\lambda\delta G_{ij}p^*_{j}\right)=\delta G_{ij}\log \left(1-\lambda p^*_{j}\right).
\end{equation}
With the above calculations, Eq.~(\ref{eq:pertuQ2}) can be written in matrix form as Eq.~\eqref{eq:dqExpress}. This completes the derivation of Eq.~\eqref{eq:dqExpress}.

\section*{Appendix B: Derivation of the incremental spreading prevalence by adding an edge}\label{sec:appendixb}
\setcounter{equation}{0}
\renewcommand\theequation{B.\arabic{equation}}
In this section we give the detailed derivation of Eq.~\eqref{strag}, i.e., the explicit formula for incremental spreading prevalence when only adding one interconnecting edge.

Since we only add one edge, then the matrix $G_{ab}=G_{ba}^{\mathrm{T}}$ can be written as an outer product
\begin{equation}
G_{ab}=uv^{\mathrm{T}},
\end{equation}
where $u$ is a vector of length $N_a$ with $u_{k}=\delta_{k,i}$ for $1\leq k\leq N_a$, and $v$ a length $N_b$ vector with $v_{k}=\delta_{k,j}$ for $1\leq k\leq N_b$. Recall $x_{ij}$ and $x_{ji}$ defined in Eq.~\eqref{eq:xDef}, then it's easy to check that
\begin{equation}
\delta X_{ab}= x_{ij} uv^{\mathrm{T}},\ \delta X_{ba}= x_{ji} vu^{\mathrm{T}}.
\end{equation}
Thus we have
\begin{equation}
\delta X_{ab} B \delta X_{ba}=x_{ij}x_{ji}B_{jj} uu^{\mathrm{T}}
\end{equation}
In other words, $\delta X_{ab} B \delta X_{ba}$ is an zero matrix expect in the $j$th element in the diagonal. The Sherman-Morrison formula says that
\begin{equation}\label{eq:Cmatrix}
\begin{split}
C&=\left(\mathbb{I}-X^0_a-x_{ij}x_{ji}B_{jj} uu^{\mathrm{T}}\right)^{-1}\\
&=A+\frac{x_{ij}x_{ji}B_{jj}A u u^{\mathrm{T}}A}{1-x_{ij}x_{ji}A_{ii}B_{jj}}.
\end{split}
\end{equation}
With this formula we can construct $\left(\mathbb{I}-X\right)^{-1}$ easily from $C$. Similarly,
\begin{equation}
\begin{split}
D&=\left(\mathbb{I}-X^0_b-x_{ij}x_{ji}A_{ii} vv^{\mathrm{T}}\right)^{-1}\\
&=B+\frac{x_{ij}x_{ji}A_{ii}B v v^{\mathrm{T}}B}{1-x_{ij}x_{ji}A_{ii}B_{jj}}.
\end{split}
\end{equation}
Again recall the definitions of $c_{ij}$ and $c_{ij}$ in Eq.~\eqref{eq:cDef}, then $y_a, y_b$ can written as
\begin{equation}
y_a= c_{ij} u,\ y_b= c_{ji} v.
\end{equation}
Combine the above computations, we arrive at the formula
\begin{equation}\label{eq:dpformula}
\begin{split}
N\delta \mathcal{P}=&\mathbf{1}^{\mathrm{T}}Cy_a+\mathbf{1}^{\mathrm{T}}C\delta X_{ab} B y_b+\\
&+\mathbf{1}^{\mathrm{T}}Dy_b+\mathbf{1}^{\mathrm{T}}D\delta X_{ba} A y_a.
\end{split}
\end{equation}
The first term on the r.h.s. of Eq.~\eqref{eq:dpformula} can be written as
\begin{equation}
\begin{split}
\mathbf{1}^{\mathrm{T}}Cy_a&=\mathbf{1}^{\mathrm{T}}Ay_a+\frac{x_{ij}x_{ji}B_{jj}}{1-x_{ij}x_{ji}A_{ii}B_{jj}} \left(\mathbf{1}^{\mathrm{T}}A\right)_i u^{\mathrm{T}}A y_a\\
&=\frac{c_{ij}}{1-x_{ij}x_{ji}A_{ii}B_{jj}} \left(\mathbf{1}^{\mathrm{T}}A\right)_i,
\end{split}
\end{equation}
where the first line is by  substituting Eq.~\eqref{eq:Cmatrix}, and second line is by using definition of $y_a$ and $u$. For the second term of r.h.s. in Eq.~\eqref{eq:dpformula},
\begin{equation}
\begin{split}
\mathbf{1}^{\mathrm{T}}C\delta X_{ab} B y_b=&c_{ji} x_{ij} B_{jj} \mathbf{1}^{\mathrm{T}}C u\\
=&\frac{c_{ji} x_{ij} B_{jj}}{1-x_{ij}x_{ji}A_{ii}B_{jj}} \left(\mathbf{1}^{\mathrm{T}}A\right)_i.
\end{split}
\end{equation}
With similar computations, we can can obtain the expressions for the rest two terms in the r.h.s. of Eq.~\eqref{eq:dpformula}, which are
\begin{equation}
\mathbf{1}^{\mathrm{T}}Dy_b=\frac{c_{ji}}{1-x_{ij}x_{ji}A_{ii}B_{jj}} \left(\mathbf{1}^{\mathrm{T}}B\right)_j,
\end{equation}
and
\begin{equation}
\mathbf{1}^{\mathrm{T}}D\delta X_{ba} A y_a=\frac{c_{ij} x_{ji} A_{ii}}{1-x_{ij}x_{ji}A_{ii}B_{jj}} \left(\mathbf{1}^{\mathrm{T}}B\right)_j.
\end{equation}
Combine the above computations we have
\begin{equation}
\begin{split}
N\delta \mathcal{P}=&\frac{c_{ij}+c_{ji} x_{ij} B_{jj}}{1-x_{ij}x_{ji}A_{ii}B_{jj}} \left(\mathbf{1}^{\mathrm{T}}A\right)_i
+\\
&+\frac{c_{ji}+c_{ij} x_{ji} A_{ii}}{1-x_{ij}x_{ji}A_{ii}B_{jj}} \left(\mathbf{1}^{\mathrm{T}}B\right)_j,
\end{split}
\end{equation}
which is Eq.~\eqref{strag}.

\end{document}